\documentclass[fleqn,usenatbib,useAMS]{mnras}

\usepackage{graphicx}	
\usepackage{amsmath}	
\usepackage{amssymb}	
\usepackage{multicol}        
\usepackage{booktabs}
\usepackage{float}
\usepackage[small,labelfont=bf,up,textfont=it,up]{caption}
\usepackage{bm}		
\usepackage{pdflscape}	

\usepackage{tabularx}
\usepackage{longtable}

\usepackage{xspace}

\newcommand{\gaia}{\emph{Gaia}\xspace}

\makeatletter
\newcommand{\@todonotes@enable}{1}
\newcommand{\@todonotes@inline}{1}
\makeatother
\input{notes.tex}

\usepackage[T1]{fontenc}
\usepackage{ae,aecompl}

\usepackage{newtxtext,newtxmath}

\title[The Effect of Dwarf Galaxies on Globular Clusters]{The Effects of Dwarf Galaxies on the Orbital Evolution of Galactic Globular Clusters}

\author[Garrow et al.]{Turner Garrow$^{1,2}$ \thanks{E-mail:  t2garrow@uwaterloo.ca (TG), webb@astro.utoronto.ca (JW)}, Jeremy J. Webb$^1$ \& Jo Bovy$^1$ \\
$^1$Department of Astronomy and Astrophysics, University of Toronto, 50 St. George Street, Toronto, ON, M5S 3H4, Canada \\
$^2$Institute for Quantum Computing and Department of Electrical and
Computer Engineering, University of Waterloo, Waterloo, N2L 3G1,
Canada \\
}

\date{}

\pubyear{2020}

\begin{document}
\label{firstpage}
\pagerange{\pageref{firstpage}--\pageref{lastpage}}
\maketitle

\begin{abstract}
We investigate the effect that dwarf galaxies have on the orbits, tidal histories, and assumed formation environment of Milky Way globular clusters. We determine the orbits of the Milky Way's 150 globular clusters in a gravitational potential both with and without dwarf galaxies. We find that the presence of a small number of satellite galaxies can affect the orbits of many of the globular clusters. Over 12 Gyr, we find that the semi-major axis and orbital eccentricity of individual clusters fluctuate with dispersions on the order of $\sim 10\%$ and $\sim 4\%$, respectively. Outer clusters are more strongly affected by dwarf galaxies than inner clusters, with their semi-major axis and orbital eccentricities fluctuating by more than  $\sim 15\%$ and $\sim 5\%$, respectively. Using detailed $N$-body simulations of select clusters, we find that altering their orbital histories can lead to different mass loss rates and structural evolution. Furthermore, we caution against using kinematics alone to identify whether a Galactic cluster formed in-situ or was accreted during a past merger event as these values are no longer conserved. The presence of dwarf galaxies causes the orbital energies and actions of individual clusters to evolve over time, spanning a wider range than that coming from random uncertainties in a cluster's proper motions and radial velocity.



\end{abstract}




\section{Introduction}\label{sec:intro}

Globular clusters (GCs) are gravitationally bound groups of stars, with ages between $11$ and $13$ Gyr \citep{Krauss65,MarinFranch09,Forbes10}, that form while their host galaxy is built up via hierarchical growth \citep{White91, Springel05}. Their subsequent evolution has been shown to be directly linked to that of their host galaxy \citep[e.g.,][]{Baumgardt03}. Hence GCs offer a window into what high redshift galaxies looked like while also having been shaped by the growth and evolution of their host. GCs are therefore ideal for studying the dynamical evolution of galaxies.

With the release of \gaia data release 2 (DR2) \citep{Gaia16,Gaia18}, the three-dimensional positions and velocities of every Galactic GC are now known \citep{Helmi18, Vasiliev19}. Knowing the orbital evolution of each cluster allows for GCs to be used as tools to study the Milky Way, which includes measuring its mass profile and identifying past merger events. For example, \citet{Eadie19} used the proper motions of 150 GCs to estimate the mass profile of the Milky Way out to 200 kpc. With respect to constraining the Milky Way's merger history, \cite{Myeong_2018} found a set of eight GCs clustered in kinematic phase space, which indicates they may have all been a part of the same progenitor galaxy that was accreted by the Milky Way in the distant past. \citet{Massari19} expanded on this work in order to identify each Galactic GC as forming in-situ or being associated with one of the Milky Way's past merger events. However the authors note that some uncertainty exists in these identifications as the kinematic properties of each of the systems overlap. Including cluster metallicity and age in the analysis can reduce the uncertainty, with \citet{Kruijssen20} further associating several Galactic GCs with the Kraken merger event suggested by \citet{Kruijssen19}.

With respect to GCs themselves, \gaia DR2 has made it easier to identify stars that are cluster members, allowing for the density profiles of individual cluster to be accurately measured and compared to dynamical models \citep{deBoer19}. Having kinematic information for a large number of cluster members allows for each clusters velocity dispersion profile to be measured, with rotation and orbital anisotropy also being observed in several GCs \citep{Bianchini18,Baumgardt18,Jindal19}. Therefore in the age of \gaia, with so much spatial and kinematic information available for GCs and individual GC stars, it is possible to directly compare observations to simulations.

Theoretical work in the field of globular cluster dynamics typically treats the Milky Way as the only significant body that influences the GCs \citep[e.g.,][]{Baumgardt03, Kruijssen09, Gieles11, Webb14}. Direct comparisons between such studies and observed GCs suggest that external factors besides the Milky Way need to be considered. For example, \citet{Webb15} found that the stellar mass functions of several Galactic GCs imply that they have lost significantly more mass than their orbits in the Milky Way would suggest \citep{Kruijssen09}. Given that clusters that show the most evidence for increased mass loss rates are spatially extended \citep{DeMarchi07, Paust10}, the mechanism behind their evolution is more likely to be external than internal. With recent studies of satellite galaxy and GC evolution in time dependent tidal fields highlighting the importance of galaxy growth and substructure \citep{Kruijssen15, Li17, Renaud17, Penarrubia19, GaravitoCamargo19, Patel20}, it is necessary to consider how individual Galactic GCs have been affected by these factors. 

The first step is to consider how the presence of dwarf satellite galaxies has affected the dynamical evolution of Galactic GCs. \citet{Gomez_2015} studied the importance of considering the Large Magellanic Cloud (LMC) in simulations of the tidal tails of the Sagittarius dwarf galaxy (Sgr) around the Milky Way. They found that when the LMC was included, tidal debris from Sgr was found to be misaligned with the present-day orbital plane of Sgr. More generally, \citet{Patel20} finds that interactions with the Milky Way and the Magellanic Clouds can affect both the orbital history and evolution of other satellites. Even more subtle changes to the external tidal field, like how the Milky Way's halo itself responds to satellite galaxies \citep{GaravitoCamargo19}, can lead to the perturbation of satellite orbits. With respect to GCs, \citet{Erkal2018} and \citet{Erkal2019} demonstrated that the Large Magellanic Cloud is capable of perturbing the orbits of the Tucana and Orphan streams, respectively. These results suggest that the presence of dwarf galaxies around the Galaxy could potentially influence the orbits of the Milky Way's other satellites as well and, in particular, that nearby dwarf galaxies will influence the orbits and subsequent evolution of Galactic globular clusters.

Finding that dwarf galaxies can strongly affect the orbital and structural evolution of GCs will significantly alter our view of how GC systems evolve. Close interactions between GCs and dwarf galaxies could lead to both episodes of mass loss due to tidal shocks and a change in the cluster's orbital path. A change in a cluster's orbit will also affect its mass loss history, such that estimates of initial mass, initial relaxation time, and the evolution of its stellar mass function will have to be revised. Furthermore, when using GC orbits to constrain the mass profile of the Milky Way one would have to factor in the presence of dwarf galaxies as well.

The purpose of this study is to determine how the presence of dwarf galaxies affects the orbital histories of Milky Way globular clusters. In Section \ref{sec:method}, we discuss how we setup the combined tidal field of the Milky Way and its satellite dwarf galaxy system in order to integrate the orbits of Galactic GCs. In Setion \ref{sec:results} we present how the orbits of Galactic GCs are changed by the presence of dwarf galaxies, mainly in terms of how their semi-major axis and orbital eccentricity differ both at the present day and as a function of time. We also consider the effects of just the Sagittarius dwarf galaxy and just the Magellanic Clouds on cluster orbits, because these are two of the more massive systems and could help rule out the importance of considering less massive dwarfs when modelling GC evolution. In Section \ref{sec:discussion}, we discuss the implications of having different orbital histories on the tidal evolution of select GCs. Through $N$-body simulations, we explore whether changes in a cluster's orbit due to interactions with dwarf galaxies can significantly affect the cluster's mass loss history. We also consider how the orbital energy and actions of each cluster are affected to see if their kinematic association with a specific merger event should be reconsidered. We summarize our findings in Section \ref{sec:conclusion}.

\section{Method}\label{sec:method}

In order to observe and quantify the effects that dwarf galaxies have on the orbital evolution of GCs, it is necessary to integrate their orbits in a Milky Way-like potential with and without the presence of dwarf galaxies. We make use of the Python galactic dynamics package \texttt{galpy}\footnote{Available at \url{http://github.com/jobovy/galpy}~.} \citep{galpy} and the orbital properties of each GC as measured by \citet{Vasiliev19}. In all cases, we assume that the Milky Way is well represented by the \texttt{MWPotential2014} mass model from \citep{galpy}.

In the base potential model, \texttt{MW}, the orbits of each cluster are solved in \texttt{MWPotential2014} only. A subset of the Milky Way's satellite galaxies are then chosen to include in the simulations based on their proximity to the Milky Way and availability of mass profile parameters. More specifically, we explore how the presence of Sagittarius, the Large and Small Magellanic Clouds, Draco, Ursa Minor and the Fornax dwarf galaxies affects the orbital evolution of GCs. Each dwarf galaxy is represented by a Hernquist mass profile, which is defined by the total mass of the dwarf and a scale radius. The values for these parameters and the appropriate references are summarized in Table \ref{dwarfs}.

\begin{table}
\centering
\begin{tabular}{@{}llcc@{}}
Galaxy      & Mass ($M_\odot$) & Scale radius (kpc) & Reference \\ \toprule
LMC         & $1.0\times10^{11}$    & 10.2 & {1,2} \\ 
SMC         & $2.6\times10^{10}$    & 3.6  & {1} \\ 
Sagittarius & $1.4\times10^{10}$    & 7.0  & {2} \\ 
Draco       & $6.3\times10^{9}$     & 7.0  & {3} \\ 
Ursa Minor  & $2.5\times10^{9}$     & 5.4  & {3} \\ 
Fornax      & $2.0\times10^{9}$     & 3.4  & {4} \\ 
\bottomrule
\end{tabular}
\caption{A summary of the dwarf galaxy properties used in the simulations, taken from (1) \citet{Besla10}, (2) \citet{Laporte18}, (3) \citet{Penarrubia08}, and (4) \citet{Goerdt06}.}
\label{dwarfs}
\end{table}

The orbits of the individual satellite galaxies are integrated backwards in time for 12 Gyr in \texttt{MWPotential2014}, given their current positions and proper motions in \citet{Helmi18}. The effects of \citet{Chandrasekhar1943} dynamical friction are included when integrating satellite galaxy orbits, using an implementation in \texttt{galpy} that is comparable to that of \citet{Petts2016}. Hence the combined potential of the Milky Way and the six satellite dwarf galaxies considered here is now known as a function of time. It is important to note that, in this study, we do not include the effects of mass loss on the satellite galaxies as the mass loss histories of dwarf galaxies are poorly constrained. Furthermore, the time evolution of the Galactic potential would also need to be considered to accurately model both effects. Given that the primary focus of this work is determining whether or not the satellites can strongly affect GC orbits, rather than trying to exactly reproduce the orbital history of individual GCs, these assumptions are valid. 

The orbit of each GC is then integrated in the combined potential of \texttt{MWPotential2014} and several different combinations of the dwarf galaxies in Table \ref{dwarfs}. We first consider cases where only the Sagittarius dwarf galaxy \texttt{SGR} or the Large and Small Magellanic Clouds \texttt{SMC/LMC} are accounted for when integrating cluster orbits. These solutions will provide a measure of how strongly the most prominent dwarf galaxies affect cluster orbits. In the final potential model \texttt{DW}, we consider GCs orbiting in the combined potential of \texttt{MWPotential2014} and all six dwarfs. For each potential model, we integrate the orbit of each cluster backwards for 12 Gyr. The four potential models are summarized in Table \ref{models}.

\begin{table}
\centering
\begin{tabular}{@{}llcc@{}}
Model Name      & Components \\ \toprule
MW         & \texttt{MWPotential2014}  \\ 
SGR         & \texttt{MWPotential2014} + Sagittarius  \\ 
SMC/LMC & \texttt{MWPotential2014} + SMC + LMC \\ 
DW       & \texttt{MWPotential2014} + SMC + LMC + \\
{} & Sagittarius + Draco + Ursa Minor + Fornax \\ 
\bottomrule
\end{tabular}
\caption{A summary of the four galaxy models used in the simulations}
\label{models}
\end{table}

\section{Results}\label{sec:results}

\begin{figure}
\includegraphics[width=0.48\textwidth]{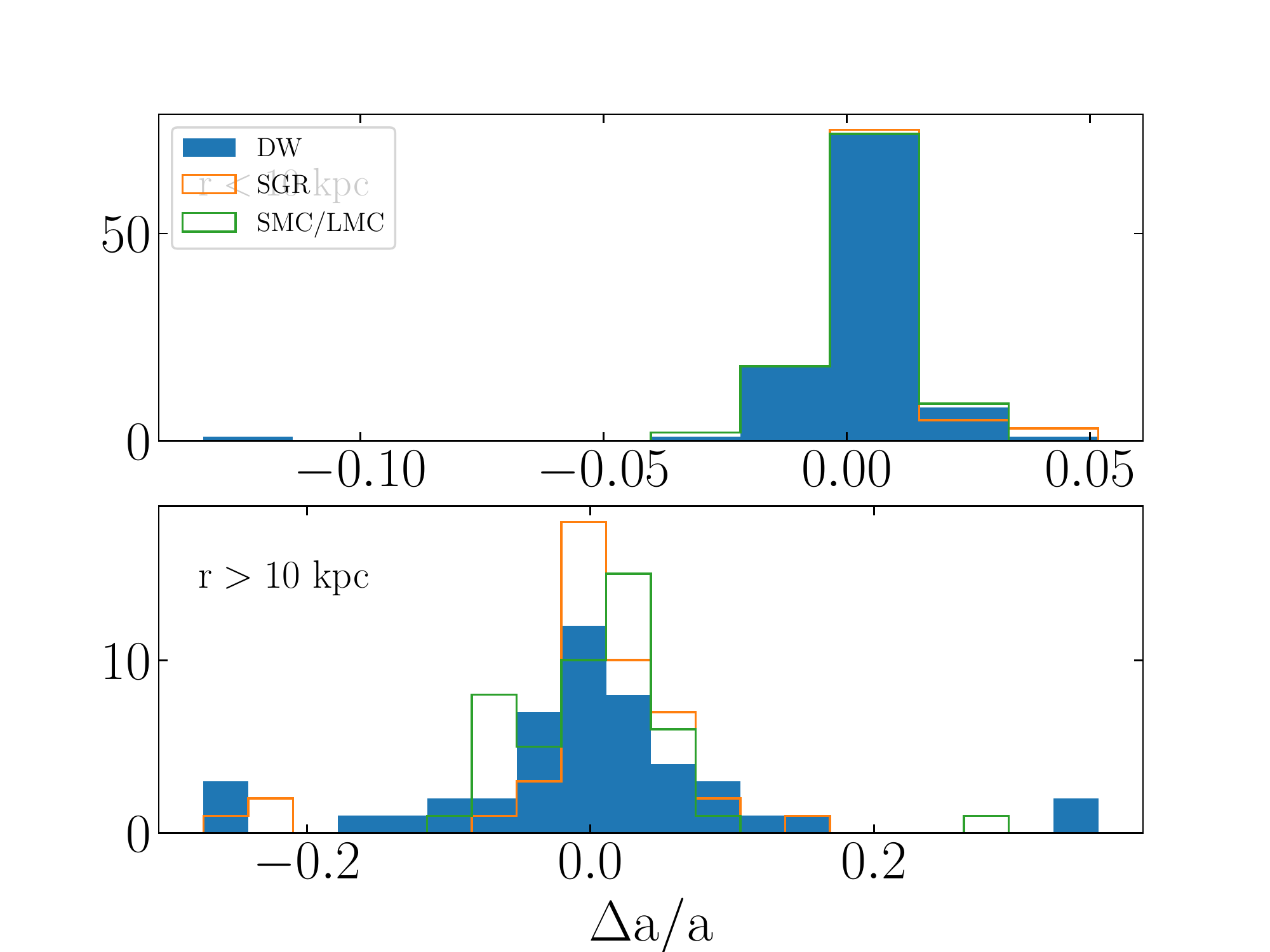}
\caption{The difference in semi-major axis of inner (top panel) and outer (bottom panel) globular clusters orbiting in the \texttt{SGR}, \texttt{SMC/LMC}, and \texttt{DW} galaxy models compared to the \texttt{MW} galaxy model, normalized by the cluster's semi-major axis in the \texttt{MW} galaxy model. Note that the scale of the $x$ axis is different in the top and bottom panels. Dwarf galaxy interactions can lead to significant changes in the semi-major axis of outer clusters.}
\label{fig:delta_a}
\end{figure}

\begin{figure}
\includegraphics[width=\columnwidth]{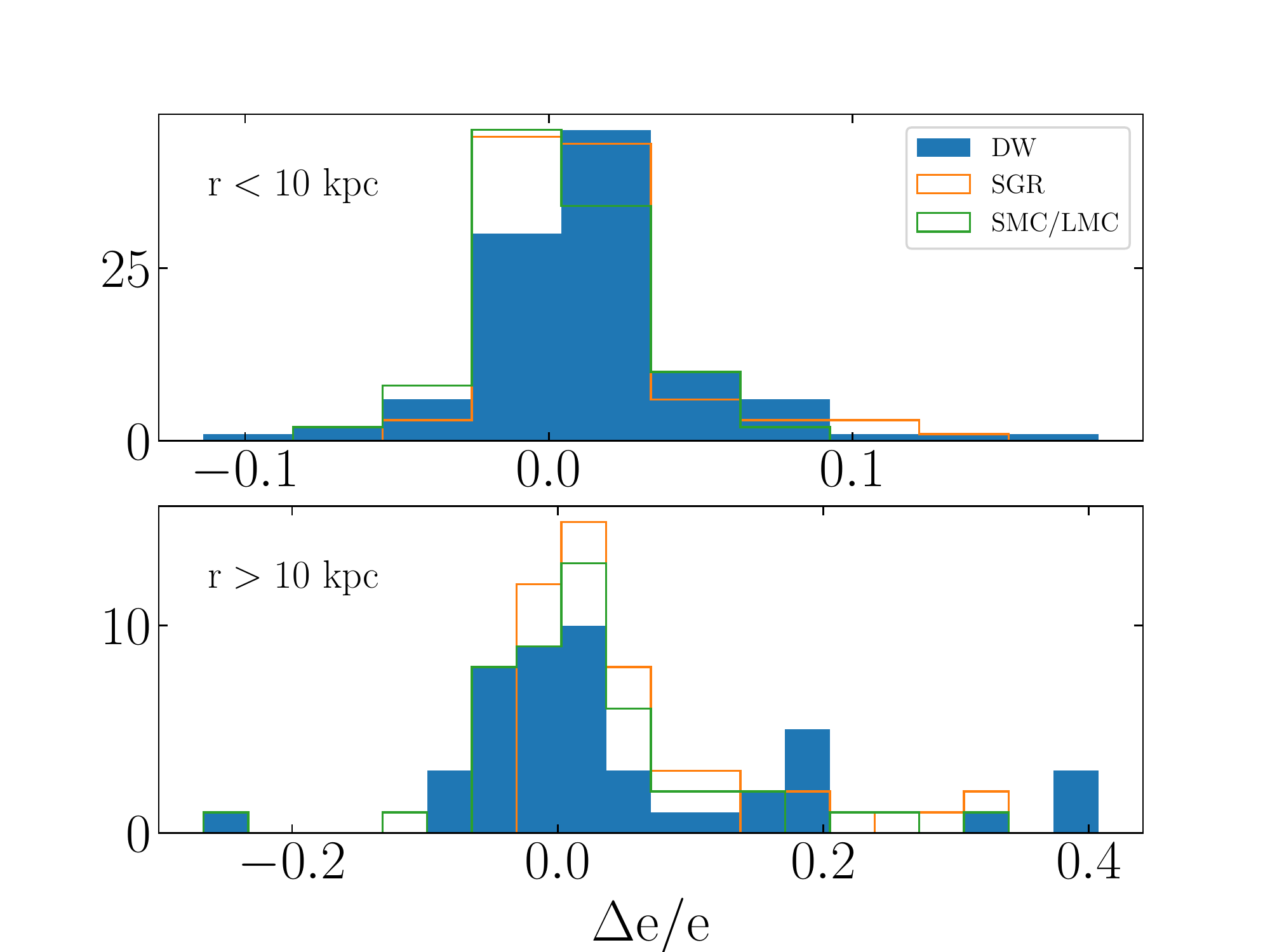}
\caption{The difference in orbital eccentricity of inner (top panel) and outer (bottom panel) globular clusters orbiting in the \texttt{SGR}, \texttt{SMC/LMC}, and \texttt{DW} galaxy models compared to the \texttt{MW} galaxy model, normalized by the cluster's eccentricity in the \texttt{MW} potential model. Note that the scale of the $x$ axis is different in the top and bottom panels. Dwarf galaxy interactions can lead to changes in the eccentricity of both inner and outer clusters.}
\label{fig:delta_e}
\end{figure}

In order to determine the effects of dwarf galaxies on the orbital evolution of GCs, we first consider how the present day semi-major axis (Figure \ref{fig:delta_a}) and orbital eccentricity (Figure \ref{fig:delta_e}) of each GC differs between the \texttt{MW} potential model and the \texttt{SGR}, \texttt{SMC/LMC}, and \texttt{DW} potential models. Table \ref{orbits} in the Appendix lists the pericentre, semi-major axis, and orbital eccentricity of each GC in each potential model. In our analysis, we consider inner ($r<10$ kpc) and outer ($r>10$ kpc) GCs separately, because outer GCs are expected to be more strongly affected by dwarf galaxy interactions since the relative strength of the dwarf galaxy potentials with respect to the Milky Way is higher. 

Figure \ref{fig:delta_a} illustrates the difference between each GC's semi-major axis in the \texttt{MW} potential model and the semi-major axis in the three dwarf galaxy models normalized by the semi-major axis in the \texttt{MW} potential model  $\Delta a /a $. Inner region clusters are minimally affected, with semi-major axes changing by less than $5\%$ in most cases. The dispersion $\sigma_a$ in $\Delta a/a$ for all three potential models is less than $2\%$. The majority of the outer region clusters experience changes to their semi-major axis that are on the order of $5\%$, however there exist several cases where the change is much larger.  The \texttt{SMC/LMC} and \texttt{SGR} potentials can lead to changes in the semi-major axis of $20\%$, with $\sigma_a=10\%$. When all six dwarf galaxies are considered in the \texttt{DW} model, the distribution in $\Delta a/a$ for outer clusters is slightly broader than the other cases ($\sigma_a = 17\%$), with select clusters experiencing changes in their semi-major axis greater than $20\%$. It is important to note that these large dispersions are driven by outliers with $\Delta a/a > 20\%$. The median absolute deviation of the \texttt{DW} model is only $5\%.$. 

Similarly, Figure \ref{fig:delta_e} illustrates the difference between each GC's orbital eccentricity in the \texttt{MW} potential model and the orbital eccentricity in the three dwarf galaxy models,normalized by the cluster's eccentricity in the \texttt{MW} potential model $\Delta e/e$. Unlike $\Delta a/a$, we find that even inner region clusters experience a change in their orbital eccentricity when the effects of dwarf galaxies are included in their orbital history. The dispersion $\sigma_e$ in $\Delta e/e$ for inner clusters in \texttt{DW} and \texttt{SGR} is $4\%$ and for \texttt{SGR} $\sigma_e$ is $3\%$. Similar to $\Delta a/a$, the orbits of outer region clusters are more strongly affected by the presence of dwarf galaxies. In the \texttt{SMC/LMC} and \texttt{SGR} models, $\sigma_e$ is approximately $4\%$ with clusters having changes in their orbital eccentricity greater than $20\%$. The \texttt{DW} models show the largest spread in $\Delta e$, with the distribution no longer appearing to be Gaussian.

\begin{figure}
\includegraphics[width=\columnwidth]{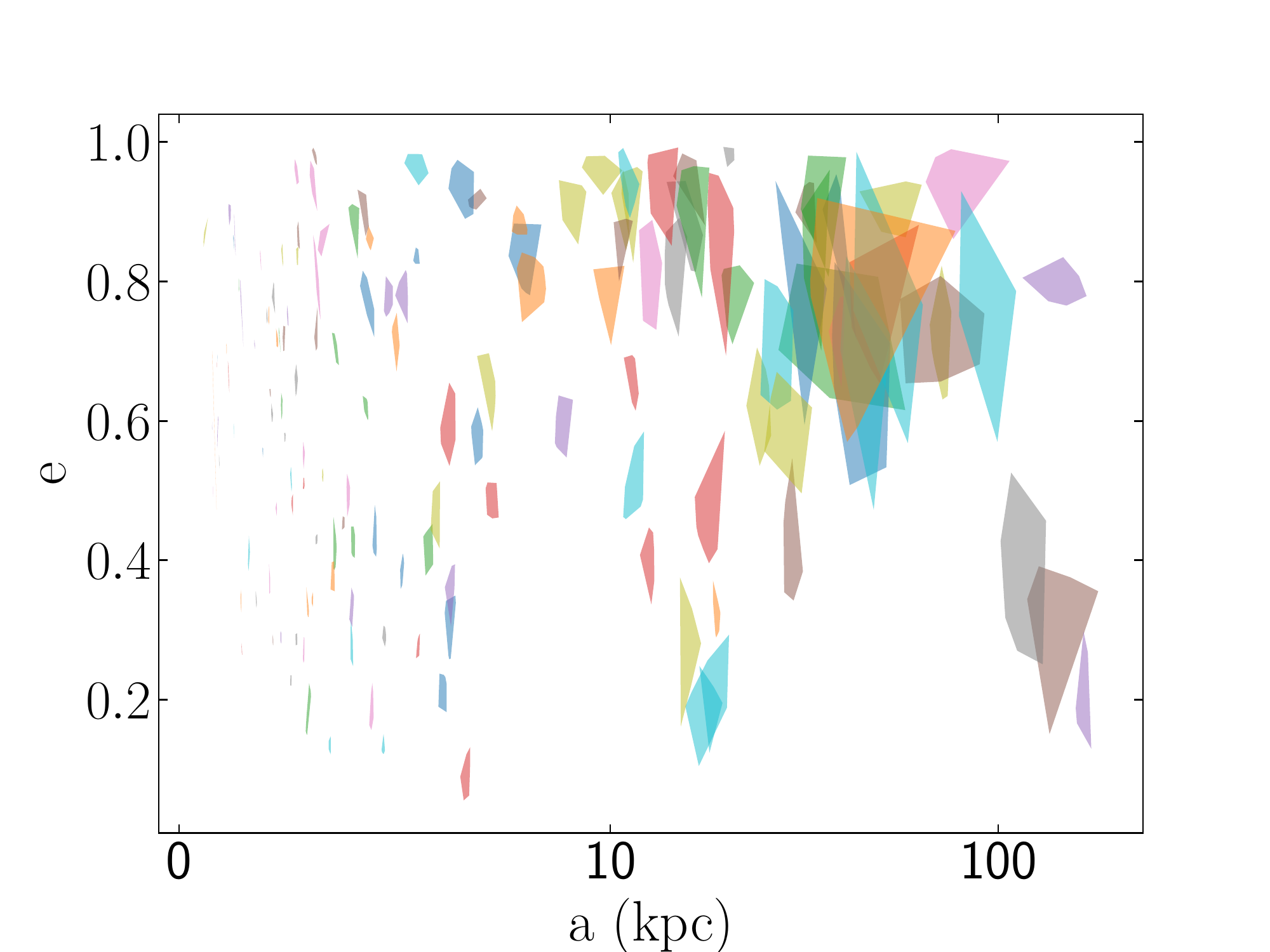}
\caption{The range of orbital eccentricities and semi-major axes---computed in the static \texttt{MW} model---that individual Galactic GCs have while orbiting in the \texttt{DW} galaxy model for 12 Gyr. Many GCs span a significant area in this plane over their past history and their present-day value is not always representative of the range of values covered.}
\label{fig:delta_range}
\end{figure}

In the \texttt{MW} model, the eccentricity and semi-major axis of the clusters do not change over the past $12$ Gyr, because the potential is static. However in the \texttt{SGR},\texttt{SMC/LMC}, and \texttt{DW} models the orbits of the clusters are expected to change over time due to repeated interactions with the dwarf galaxies. Therefore, we take the position and velocity of the globular clusters integrated in the \texttt{DW} model at $1$ Gyr timesteps and numerically determine their semi-major axis and orbital eccentricity in the \texttt{MW} potential through orbit integration to explore the time evolution of these two parameters; that is, this calculation shows how the orbital parameters change because of perturbations from dwarf galaxies when computed in a static model. The range in orbital properties of each GC over the past $12$ Gyr is shown in Figure \ref{fig:delta_range}, with each shaded area representing the range in semi-major axis and orbital eccentricity parameter space that each GC covers. 

Figure \ref{fig:delta_range} clearly demonstrates that the presence of a subset of Milky Way satellite galaxies has a significant effect on the orbits of the globular clusters as a function of time. Outer clusters are, as expected, more strongly affected by the presence of dwarf galaxies as they span a wider range in semi-major axis and orbital eccentricity than inner clusters. The orbits of select inner region clusters also appear to be affected by the presence of dwarfs.Note that by only calculating values at $1$ Gyr timesteps we are not illustrating the full range in eccentricity and semi-major axis that each cluster experiences, but merely a representative sampling. Hence the area filled by each cluster in Figure \ref{fig:delta_range} would be larger with higher time-resolution.

\section{Discussion}\label{sec:discussion}

The orbit integrations above of each Galactic GC in tidal fields with and without satellite dwarf galaxies reveal that both the present day and past orbits of GCs can be affected by interactions with dwarfs. The interactions can result in a change in both the GC's semi-major axis and orbital eccentricity. Such changes can potentially affect the dynamical evolution of a GC, because cluster evolution is strongly coupled to the properties of its host galaxy. Furthermore, if a cluster's orbit was significantly different in the past than it is today, using its current orbital properties to constrain its origin (in-situ formation or accretion via a merger event) becomes difficult. We discuss both of these factors in the following sections.

\subsection{Dynamical Evolution}

The presence of dwarf galaxies changes the tidal history of a GC in two ways. First, because dwarf galaxies can change a cluster's orbit, they also change the tidal field that a GC experiences. Hence the tidal field experienced by individual globular clusters in the past may be stronger or weaker than what they experience on their current orbit. Secondly, the gravitational force from dwarf galaxies can also itself be directly responsible for tidally stripping stars. However given that the tidal force scales as $\sim 1/{r^3}$, this effect is only significant if a globular cluster has a close encounter with a dwarf galaxy.

In order to explore these two channels, we simulate the evolution of star clusters for a few select Galactic GC orbits using \texttt{gyrfalcON}, a force calculation algorithm and $N$-body code within NEMO \citep{teuben95} that is capable of including an external tidal field \citep{dehnen00,dehnen02}. To select which GCs to simulate, we separate GCs into four subgroups based on semi major axis (a < 10 kpc, 10 < a < 20 kpc, 20 < a < 50 kpc, and 50 < a < 200 kpc) and then identify the GC within each subgroup that spans the largest area in Figure \ref{fig:delta_range}. This selection criteria allows us to identify clusters over a range of tidal field strengths that have had their orbit affected by the presence of dwarf galaxies the most. In order of increasing semi-major axis, these clusters are NGC 6528, IC 1257, Pal 2, and Pal 4. Their internal evolution is modelled with \texttt{gyrfalcON} assuming stars have a softening length of 1.5 pc, an opening angle of 0.6, and a maximum timestep size of 3.9 Myr. The tidal field of the Milky Way is analytic and set equal to \texttt{MWPotential2014}, while the dwarf galaxies are included as heavily softened particles. Given that none of the \texttt{GYRFALCON} softening mechanisms correspond to a Hernquist sphere, we instead assume each of the six dwarf galaxies in Table \ref{dwarfs} are Plummer spheres with the same mass and scale radius. 

The initial position and velocity of each GC and the dwarf galaxies in \texttt{MWPotential2014} are set to what their values must be 12 Gyr ago in order to reach their present day locations in the Galaxy. Each GC consists of 10,000 stars and has a total mass of $2 \times 10^5\,M_{\odot}$. The half mass radius $r_m$ of the GC is set so that it is tidally filling (ratio of the half mass radius and the tidal radius $r_t$ of $r_m/r_t = 0.145$) at its initial pericentre in the \texttt{DW} model \citep{Henon61,Henon65}. Setting the clusters to be filling at pericentre ensures that they are sensitive to changes in the external tidal field. For comparison purposes, the exact same initial cluster is simulated in the \texttt{MWPotential2014} tidal field only, with its initial position and velocity in the galaxy set so that it too will end up at the cluster's current position in the Milky Way. A direct comparison between simulations with and without dwarfs allows us to determine how strongly a cluster's evolution has been affected by the presence of dwarf galaxies. For each GC, five different simulations in the \texttt{MWPotential2014} tidal field are run where the present-day proper motion and radial velocity of the cluster are randomly chosen within the uncertainties quoted by \citet{Vasiliev19}. This approach allows for the dynamical evolution of GCs in simulations with dwarf galaxies to be compared to the range of allowable orbital histories GCs can have due to the uncertainty in their kinematic properties. 

\begin{figure}
\includegraphics[width=\columnwidth]{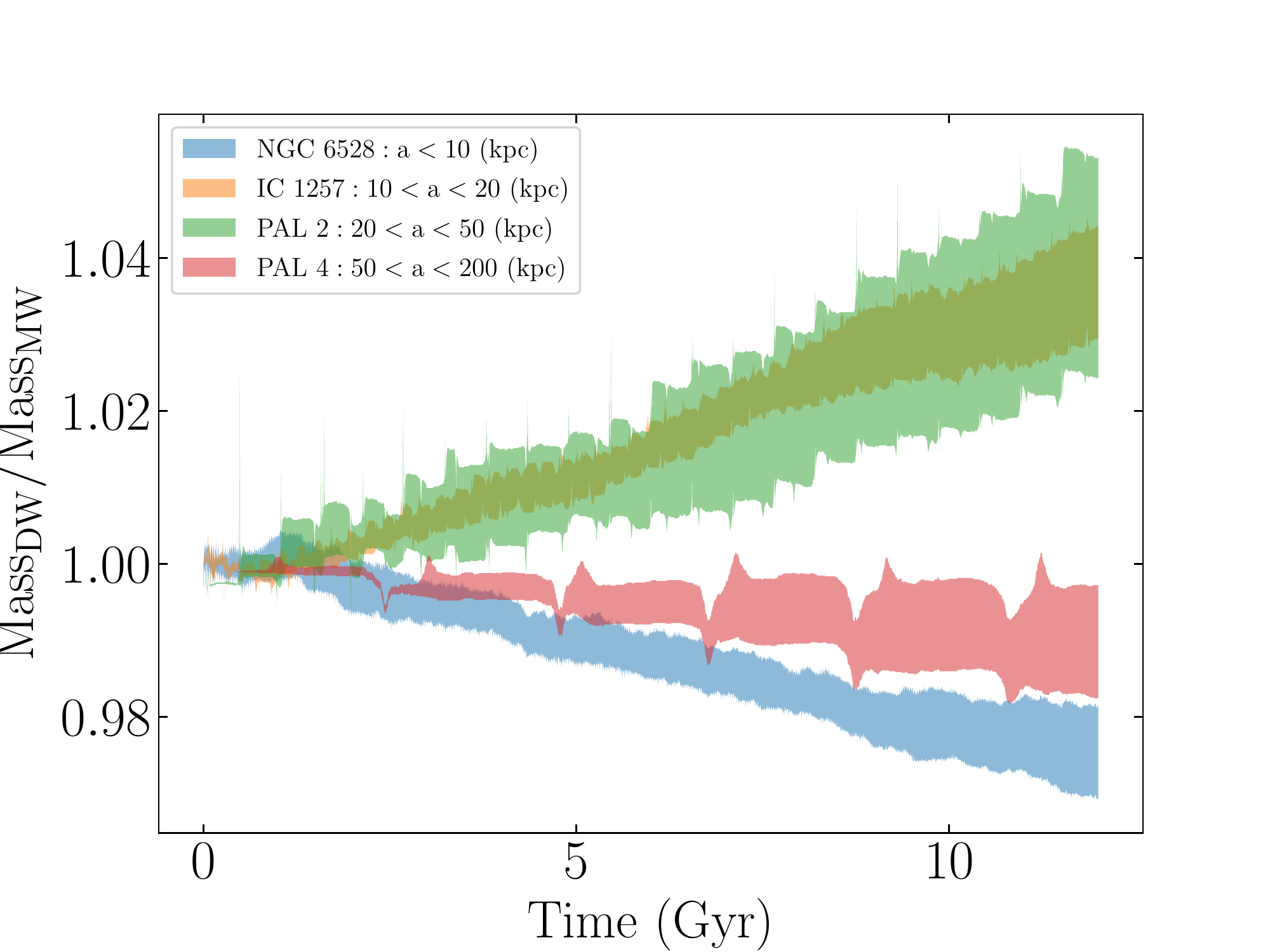}
\caption{The mass loss histories of four globular clusters selected in four semi-major axis ranges to have large changes in their orbit due to interactions with dwarf galaxies. Each shaded region represents the mass evolution of a cluster in the \texttt{DW} galaxy model normalized by the mass evolution of five clusters simulated in the \texttt{MW} that have had their initial proper motions and radial velocity randomly drawn from within their quoted uncertainties. The masses of the clusters were found by counting stars within two tidal radii of the densest region of the cluster. Globular clusters can lose a few percent more or less mass due to the effect of dwarf galaxies.}
\label{mass_loss}
\end{figure}

\begin{figure}
\includegraphics[width=\columnwidth]{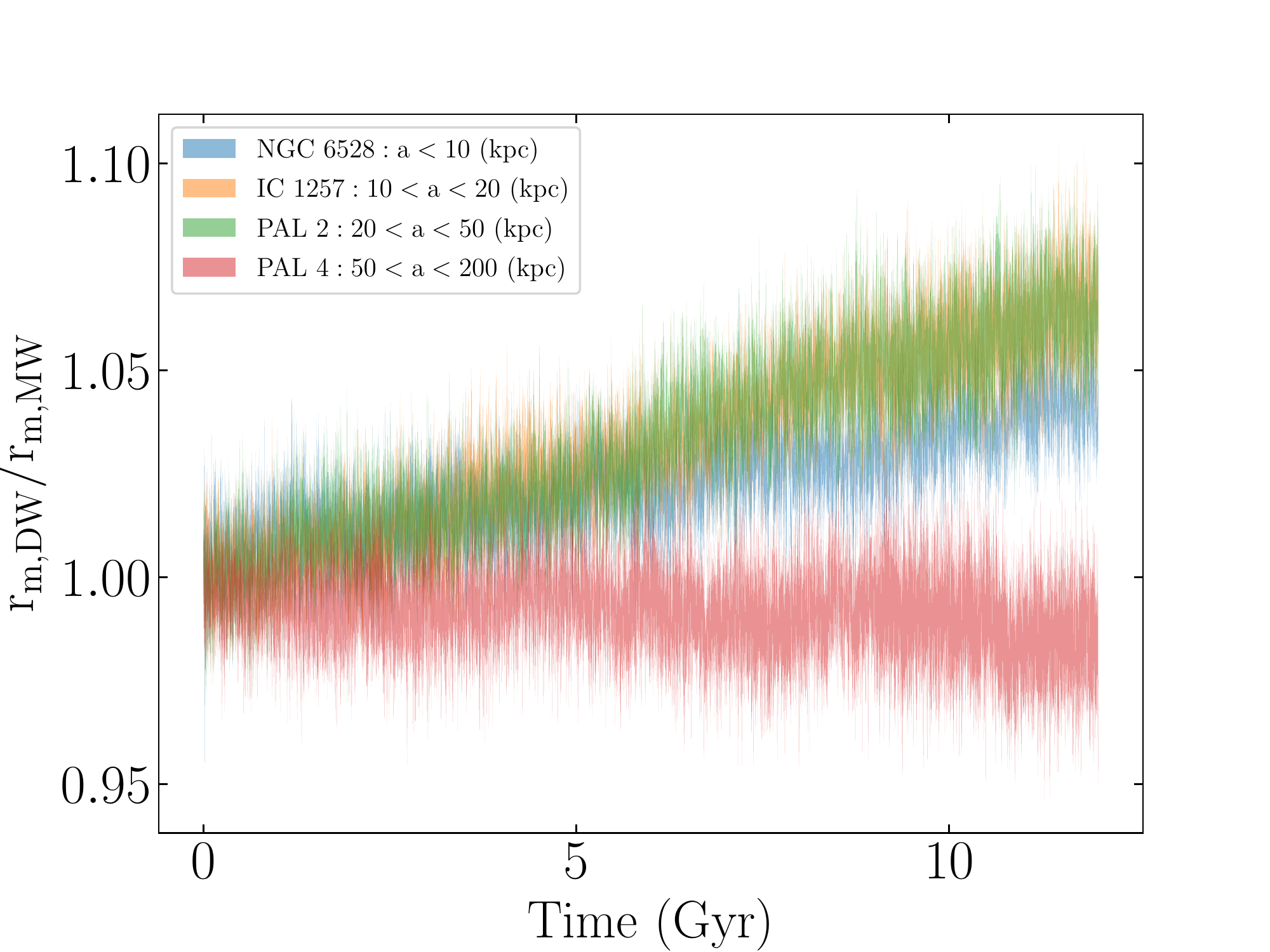}
\caption{Same as Figure \ref{mass_loss}, but for cluster half-mass radii. Globular clusters in a stronger tidal field ($a < 50$ kpc) are all larger due to the effect of dwarf galaxies.}
\label{r_hm}
\end{figure}

\begin{figure*}
\includegraphics[width=\textwidth]{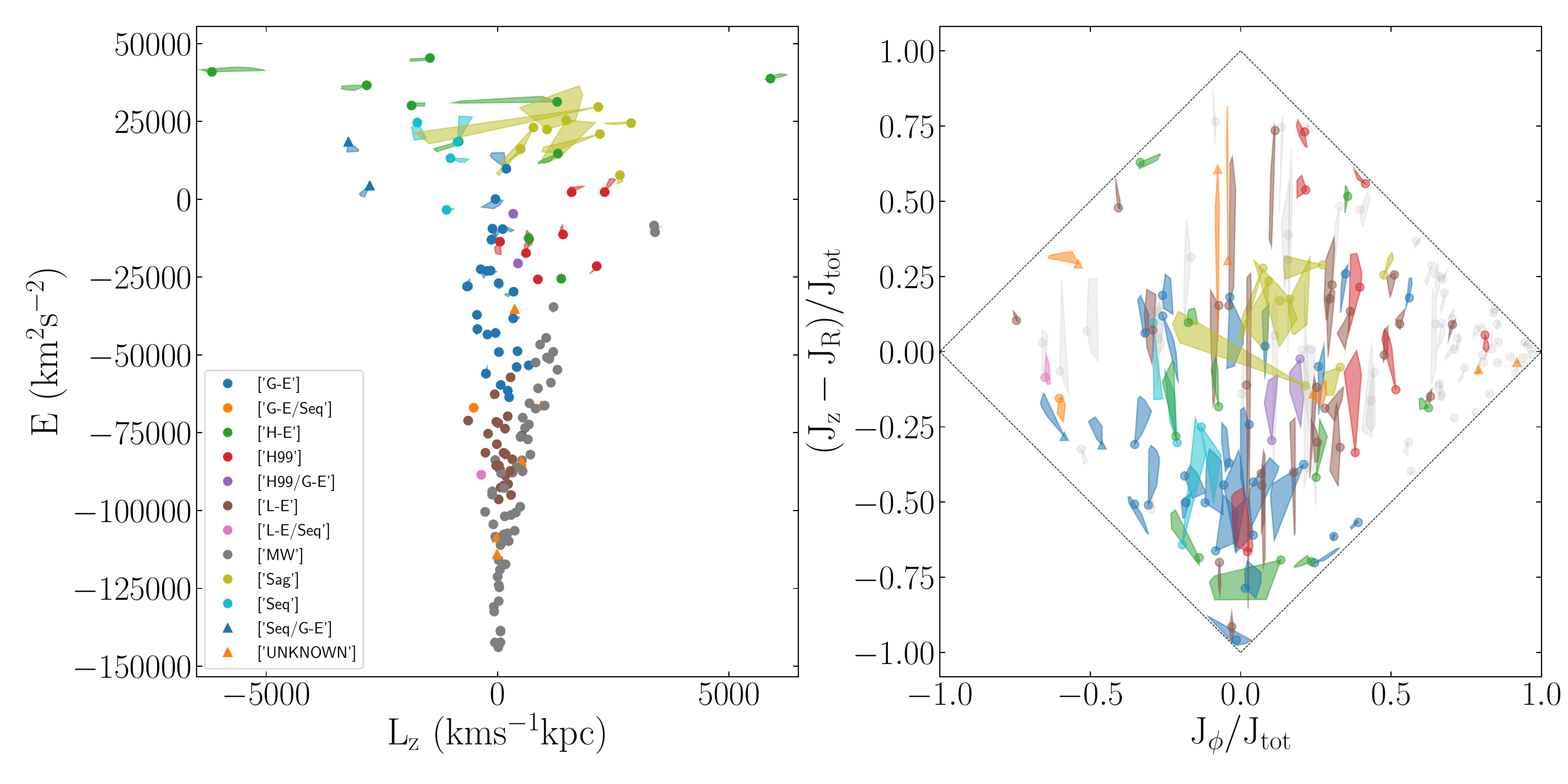}
\caption{Effect of dwarf-galaxy perturbations on integrals of the motion commonly used to identify the origin of Milky Way globular clusters. Left Panel: Range of orbital energies E and angular momenta $L_z$ reached by Galactic globular clusters evolving in the \texttt{DW} galaxy model.  Clusters are colour coded based on the work of \citet{Massari19}, who uses this parameter space to designate clusters as being associated with either \gaia-Enceladus (G-E), the Sagitarius dwarf galaxy (Sgr), the Helmi stream progenitor (H99), the Sequoia galaxy (Seq), the Low-Energy group (L-E), the High-Energy group (H-E), or the Milky Way (MW). (Note E1, which is not associated with any group, is not featured in this diagram due to its large angular momentum $L_z < -12000 \ km \ s^{-1} \ kpc$. Right Panel: Difference between the vertical and radial actions compared to the azimuthal action (both normalized by the total action), a parameter space comparable to that used by \citet{Myeong19} to study kinematically related globular clusters. Perturbations due to dwarf galaxies can lead to clusters evolving throughout both parameter spaces over the course of their lifetime.}
\label{kinplot}
\end{figure*}

Figure \ref{mass_loss} illustrates the ratio of mass of all stars within the tidal radius (also referred to as the Jacobi radius) of the cluster orbiting in the \texttt{DW} potential normalized by the mass within the tidal radius of the clusters orbiting in \texttt{MWPotential2014}---the range results from the five different present-day positions and velocities that we simulate in \texttt{MWPotential2014}. Note that the tidal radius is first calculated assuming all the stars are bound to the cluster, and then again using only the mass of stars with radii within the initial calculation. First taking into consideration the mass evolution of the innermost cluster, NGC 6528, our simulations show that the cluster loses between $2-3\%$ more mass in the galaxy model with dwarf galaxies than without. Hence its orbit must have originated in a strong tidal field before interactions with dwarf galaxies caused it to migrate outwards. The opposite is true for IC 1257 and Pal 2, both of which lose less mass in the galaxy model with dwarf galaxies. Finally the dynamical evolution of Pal 4, which has a semi-major axis near 70 kpc, is negligibly affected despite the dwarf galaxies causing its orbit to migrate significantly due to the external tidal field being so weak.

Taking into consideration the structural evolution of each cluster, Figure \ref{r_hm} shows the ratio of the half-mass radius of all stars within the tidal radius of the cluster orbiting in the \texttt{DW} potential normalized by the half mass radii of all stars within the tidal radius of the clusters orbiting in \texttt{MWPotential2014}---the range again results from the five different present-day positions and velocities that we simulate in \texttt{MWPotential2014}. We find that the size of NGC 6528, IC 1257, and Pal 2 all increase by a factor of almost $5\%$ due to the presence of dwarf galaxies. The expansion in NGC 6528 is likely due to its tidal radius increasing as it moves from a stronger to a weaker tidal field. Conversely, IC 1257 and Pal 2 are able to initially expand to a larger size due to starting in a weaker tidal field before moving to its final orbit. Similar to Figure \ref{mass_loss}, the evolution of Pal 4 is primarily unaffected by any changes in its orbit due to dwarf galaxies because the background potential is weak regardless of its exact orbit.

\subsection{Origin and Ancient Merger Associations}

Finally, we consider the kinematic properties of each Galactic GC and how their evolution due to the presence of dwarf galaxies may influence their implied origin. The left panel of Figure \ref{kinplot} illustrates the spreads in orbital energy and angular momentum that each GC experiences over the course of its evolution in the \texttt{DW} model. This parameter space was used by \citet{Massari19} to label clusters as being members of either \gaia-Enceladus, the Sagitarius dwarf galaxy, the Helmi stream progenitor, the Sequoia galaxy, the Low-Energy group, the High-Energy group, or as being formed in-situ in the Milky Way. Clusters in the Low-Energy group have been further estimated to be part of the Milky Way's most ancient merger event, which \citet{Kruijssen20} has dubbed Kraken. It should be noted that clusters in the High-Energy group are not believed to share a common origin with any clusters, including themselves. These designations are noted in the legend, where in some cases two possible origins are listed and in others no origin was identified (UNKNOWN). More recently, \citet{Naidu20} used this parameter space to identify three halo structures in addition to \citet{Massari19}. The authors further conclude that the stellar halo is primarily made up of substructure.

The left panel of Figure \ref{kinplot} illustrates that only the energy and angular momentum of clusters associated with the Sagittarius Dwarf and the High-Energy group, and to a lesser extent clusters associated with the Helmi Stream and Sequoia Galaxy, are affected when accounting for the effects of dwarf galaxies on their orbital evolution. While not shown, in most cases the median uncertainty in both energy and angular momentum is comparable to the size of the data points. The High-Energy clusters and a select few Low-Energy clusters are the exceptions to this statement. Hence, the spread in orbital energy and angular momentum found for these clusters due to the presence of dwarf galaxies is greater than or equal to the random error that can be attributed to the uncertainties in their proper motions and radial velocities. It is perhaps not surprising that the orbits of clusters associated with Sagittarius are affected by the dwarf itself, with a detailed model for the time-evolution of the Sagittarius dwarf being necessary before it can be determined whether the kinematic properties of these clusters are consistent with once being members of Sagittarius or not. The effect of dwarf galaxies on select clusters associated with the Helmi stream and the Sequoia Galaxy could lead to them being incorrectly identified as associated clusters using kinematics alone. In the orbital energy vs. angular momentum parameter space, accounting for dwarf galaxies leads to some overlap between clusters associated with the Helmi stream and clusters believed to have formed in-situ within the Milky Way. Additionally, given how close the Sequoia galaxy and \gaia-Enceladus are in this parameter space, accounting for the effects of dwarf galaxies makes the two populations difficult to separate based on kinematics alone. A closer look at how the orbits of the stellar debris associated with each system's progenitor is affected by dwarf galaxies is required to determine whether these clusters can still be associated with a given progenitor or not. However, given that the clusters do not spread out uniformly in this parameter space, it is likely that stellar debris will do the same and the uncertainty in associations will grow as different systems being to overlap like the Sequoia galaxy and \gaia-Enceladus clusters. Similarly, extending the above conclusions to the recent work of \citet{Naidu20}, the presence of dwarf galaxies will also affect the ability to separate the stellar halo into its individual substructure components.

The right panel of Figure \ref{kinplot} illustrates the spread in orbital actions ($J_R$,$J_{\phi}$,$J_z$) that GCs have due to the presence of dwarf galaxies.  More specifically, we plot $(J_z-J_R)/J_\mathrm{tot}$ as a function of $J_{\phi}/J_\mathrm{tot}$, with colour coding remaining the same as the left panel. \citet{Myeong19} uses a similar parameter space to study the orbital characteristics of groups associated with a given progenitor.


Including the presence of dwarf galaxies when integrating the orbits of Galactic GCs results in them having a wider range of actions over the course of their lifetime. Similar to the left panel of Figure \ref{kinplot}, GCs associated with the Sagittarius Dwarf show evolution along both axis. Unlike the left panel of Figure \ref{kinplot}, GCs associated with \gaia-Enceladus and the Sequoia galaxy also evolve within this parameter space. Furthermore, GCs believed to be part of the Helmi stream have a wide range in $(J_z-J_R)/J_{tot}$ and do not appear to similar orbital characteristics. Hence taking into consideration the effects of satellite dwarf galaxies on the orbital evolution of Galactic GCs makes it difficult to use energies and actions alone to identify a cluster's association with an accretion event. 

Similar to the left panel of Figure \ref{kinplot}, the spread in orbit actions caused by the presence of dwarf galaxies is again larger than the error associated with the uncertainties in each cluster's proper motions and radial velocity (with a few similar exceptions). Hence the systematic uncertainty in each cluster's orbit attributed to the inclusion or exclusion of satellite galaxies (only 6 of which are included in this study) dominates over uncertainties associated with how well their kinematic properties are measured.

\section{Conclusions}\label{sec:conclusion}

The presence of a subset of the Milky Way's satellite galaxies has a number of effects on the population of Galactic globular clusters. Dwarf galaxies primarily affect the orbits of the Milky Way's globular clusters, with several clusters having a range of semi-major axis and orbital eccentricities over the course of their lifetime. There doesn't appear to be a significant trend in the change in any of the orbital parameters of the population of globular clusters, including apocentre, pericentre and eccentricity. However we do find evidence that the SMC and LMC are the dominant source of orbital perturbations.

The change in individual cluster orbits can have a subsequent effect on the mass loss rate and structure of a GC. Clusters with orbits that were closer to the Galactic centre before satellite galaxy interactions caused the cluster to migrate outwards lose more mass earlier in their lifetime than one would estimate assuming their orbit has remained unchanged. Similar effects are observed for clusters that have had their eccentricities grow due to an increase in their apocentre. The opposite is true for clusters that have migrated inwards or had their eccentricity grow due to a decrease in their pericentre. For the four sample clusters studied here, evolving model GCs in tidal fields with and without satellite galaxies resulted in mass loss rates varying by $\pm 3\%$ and cluster half-mass radii varying by over $5\%$. These variations are larger than any difference from uncertainties in each cluster's measured proper motions and radial velocity.

We also explore the range of orbital energies and actions that each cluster has over the course of its orbital history. Interactions with satellite galaxies can lead to the orbital energies and actions of clusters changing over time, adding a degree of uncertainty in using a cluster's kinematic properties to associate it with one of the Milky Way's past accretion events. This systematic uncertainty is often larger than the random uncertainty that can be attributed to measurements the cluster's velocity, and highlights the importance of also using chemical abundances and ages to constrain the Milky Way's accretion history \citep{2019MNRAS.482.3426M,Kruijssen19, Kruijssen20}.

There are a few important limitations that arise in our orbital integration technique that are worth noting. First, our method does not include the orbital evolution of the dwarf galaxies themselves. Interactions between satellites are also not considered, which \citet{Patel20} finds can alter the orbital history of individual satellites. Furthermore, we do not include the mass and structural evolution of the Milky Way or the dwarf galaxies throughout the past $12$ Gyr. While the mass of the Milky Way grows over time, satellites lose mass over time through tidal stripping from interactions with the Milky Way. Hence the mass of each dwarf galaxy was larger in the past, while the mass of the Milky Way was smaller. Each of these assumptions could be remedied by including a mass growth history for the Milky Way and a mass loss history for each dwarf. However, at present most satellite galaxies are lacking models for their mass and structural evolution. A full model for the mass, structural, and orbital evolution of the Milky Way's satellite population is therefore required in order to fully understand how each Galactic GC is affected by their presence. Other forms of substructure, such as dark matter subhalos, giant molecular clouds, the Galactic bar, and spiral arms are also worth considering.

Ultimately, what this study demonstrates, is that the orbital properties of GCs are not constant in time. Interactions with satellite galaxies, amongst other forms of substructure, can result in the orbital eccentricity, semi-major axis, and other integrals of the motion of a cluster evolving over time. Given how strongly a cluster's evolution is tied to the tidal field of its host galaxy, orbital evolution could potentially have strong implications on a cluster's mass-loss and structural history. It is therefore necessary to first consider possible sources that can alter the orbits of GCs before their dynamical clocks can be rewound to study their properties at formation and the evolution of their host galaxy.

\section*{Acknowledgements}
The authors would like to thank Gurtina Besla and Nicolas Garavito for valuable feedback regarding the project and James Lane for help with the implementation of the gravitational potential of orbiting satellite galaxies and generating Figure \ref{kinplot}. The authors would also like to thank the anonymous referee for constructive feedback on the manuscript. JB acknowledges financial support from NSERC (funding reference numbers RGPIN-2015-05235 \& RGPIN-2020-04712) and an Ontario Early Researcher Award (ER16-12-061).

\section*{Data Availability}

The data underlying this article are available in the public domain \citep{Goerdt06, Besla10, Penarrubia08, Laporte18, Vasiliev19} and in the article itself. 





\bibliographystyle{mnras}
\bibliography{refs} 

\appendix
\section{Orbital parameter changes in different galaxy models for all Milky Way globular clusters}

See Table \ref{orbits}.

\onecolumn
\begin{center}
\begin{longtable}{@{}lcccccccccccc@{}}
    \hline
    Galaxy Model: & {} & {MW} & {} & {} & {SGR} & {} & {} & {SMC/LMC} & {} & {} & {DW} & {} \\
    \hline
    \hline
    ID & $r_p$ & $a$ & $e$ & $r_p$ & $a$ & $e$ & $r_p$ & $a$ & $e$ \\
    \hline
    NGC104 & 5.90 & 6.67 & 0.12 & 5.80 & 6.65 & 0.13 & 6.07 & 6.76 & 0.10 & 6.06 & 6.75 & 0.10 \\
    NGC288 & 2.16 & 7.18 & 0.70 & 2.30 & 7.25 & 0.68 & 2.10 & 7.15 & 0.71 & 2.18 & 7.19 & 0.70 \\
    NGC362 & 0.28 & 6.39 & 0.96 & 0.41 & 6.47 & 0.94 & 0.21 & 6.24 & 0.97 & 0.24 & 6.32 & 0.96 \\
    Whiting1 & 21.32 & 48.01 & 0.56 & 14.94 & 44.33 & 0.66 & 16.46 & 44.14 & 0.63 & 14.02 & 40.44 & 0.65 \\
    NGC1261 & 1.23 & 11.30 & 0.89 & 1.24 & 12.20 & 0.90 & 0.80 & 11.16 & 0.93 & 0.82 & 11.15 & 0.93 \\
    Pal1 & 15.69 & 18.00 & 0.13 & 15.13 & 18.26 & 0.17 & 15.67 & 18.59 & 0.16 & 14.79 & 18.05 & 0.18 \\
    E1 & 107.72 & 169.69 & 0.37 & 108.16 & 168.60 & 0.36 & 124.62 & 171.12 & 0.27 & 124.62 & 170.20 & 0.27 \\
    Eridanus & 16.44 & 92.23 & 0.82 & 16.85 & 91.65 & 0.82 & 21.34 & 93.64 & 0.77 & 21.90 & 89.70 & 0.76 \\
    Pal2 & 0.17 & 21.03 & 0.99 & 0.32 & 21.04 & 0.98 & 0.26 & 21.53 & 0.99 & 0.19 & 20.85 & 0.99 \\
    NGC1851 & 0.40 & 10.93 & 0.96 & 0.38 & 11.14 & 0.97 & 1.00 & 11.03 & 0.91 & 0.49 & 10.82 & 0.95 \\
    NGC1904 & 0.14 & 9.77 & 0.99 & 0.19 & 10.24 & 0.98 & 0.47 & 9.63 & 0.95 & 0.36 & 9.68 & 0.96 \\
    NGC2298 & 2.13 & 10.14 & 0.79 & 2.28 & 10.34 & 0.78 & 2.07 & 10.09 & 0.79 & 2.23 & 10.16 & 0.78 \\
    NGC2419 & 16.43 & 53.96 & 0.70 & 14.70 & 53.07 & 0.72 & 15.58 & 55.58 & 0.72 & 13.71 & 53.85 & 0.75 \\
    Pyxis & 25.10 & 157.57 & 0.84 & 24.70 & 155.63 & 0.84 & 31.64 & 156.00 & 0.80 & 31.38 & 150.72 & 0.79 \\
    NGC2808 & 0.79 & 7.73 & 0.90 & 0.78 & 7.78 & 0.90 & 0.83 & 7.78 & 0.89 & 0.90 & 7.89 & 0.89 \\
    E3 & 9.05 & 13.14 & 0.31 & 9.02 & 13.24 & 0.32 & 8.16 & 12.83 & 0.36 & 8.17 & 12.99 & 0.37 \\
    Pal3 & 75.35 & 110.18 & 0.32 & 75.25 & 111.61 & 0.33 & 78.87 & 112.42 & 0.30 & 81.44 & 113.99 & 0.29 \\
    NGC3201 & 8.43 & 26.45 & 0.68 & 7.36 & 25.48 & 0.71 & 8.85 & 25.98 & 0.66 & 8.85 & 25.52 & 0.65 \\
    Pal4 & 20.54 & 68.74 & 0.70 & 20.57 & 68.78 & 0.70 & 19.40 & 71.07 & 0.73 & 20.14 & 70.89 & 0.72 \\
    Crater & 103.98 & 125.53 & 0.17 & 102.11 & 124.56 & 0.18 & 100.35 & 127.83 & 0.22 & 103.36 & 135.49 & 0.24 \\
    NGC4147 & 0.86 & 14.40 & 0.94 & 0.67 & 15.18 & 0.96 & 0.66 & 15.24 & 0.96 & 0.63 & 15.63 & 0.96 \\
    NGC4372 & 3.03 & 5.10 & 0.41 & 3.11 & 5.16 & 0.40 & 3.12 & 5.12 & 0.39 & 3.14 & 5.13 & 0.39 \\
    Rup106 & 5.30 & 24.14 & 0.78 & 5.42 & 24.67 & 0.78 & 6.70 & 25.42 & 0.74 & 7.04 & 25.63 & 0.73 \\
    NGC4590 & 8.94 & 24.92 & 0.64 & 8.73 & 24.91 & 0.65 & 10.21 & 26.22 & 0.61 & 10.21 & 27.85 & 0.63 \\
    NGC4833 & 0.62 & 4.40 & 0.86 & 0.64 & 4.45 & 0.86 & 0.60 & 4.36 & 0.86 & 0.60 & 4.41 & 0.86 \\
    NGC5024 & 9.47 & 16.45 & 0.42 & 8.89 & 17.06 & 0.48 & 9.77 & 17.17 & 0.43 & 9.60 & 17.13 & 0.44 \\
    NGC5053 & 11.66 & 14.77 & 0.21 & 10.65 & 14.77 & 0.28 & 11.28 & 15.60 & 0.28 & 11.29 & 15.93 & 0.29 \\
    NGC5139 & 1.68 & 4.35 & 0.61 & 1.71 & 4.38 & 0.61 & 1.69 & 4.34 & 0.61 & 1.63 & 4.33 & 0.62 \\
    NGC5272 & 5.28 & 11.31 & 0.53 & 4.99 & 11.17 & 0.55 & 5.36 & 11.65 & 0.54 & 5.37 & 11.62 & 0.54 \\
    NGC5286 & 1.14 & 7.92 & 0.86 & 1.05 & 7.86 & 0.87 & 1.21 & 7.99 & 0.85 & 1.24 & 7.99 & 0.84 \\
    NGC5466 & 6.90 & 41.28 & 0.83 & 7.05 & 43.25 & 0.84 & 5.65 & 52.37 & 0.89 & 6.80 & 56.07 & 0.88 \\
    NGC5634 & 3.01 & 12.36 & 0.76 & 2.61 & 12.43 & 0.79 & 2.57 & 12.60 & 0.80 & 2.51 & 12.74 & 0.80 \\
    NGC5694 & 2.53 & 40.05 & 0.94 & 3.04 & 42.19 & 0.93 & 2.15 & 39.56 & 0.95 & 2.59 & 39.14 & 0.93 \\
    IC4499 & 7.42 & 20.87 & 0.64 & 6.54 & 20.98 & 0.69 & 5.46 & 20.56 & 0.73 & 5.00 & 20.58 & 0.76 \\
    NGC5824 & 17.17 & 30.59 & 0.44 & 16.40 & 30.01 & 0.45 & 16.81 & 29.82 & 0.44 & 17.45 & 29.72 & 0.41 \\
    Pal5 & 15.02 & 17.54 & 0.14 & 15.01 & 18.38 & 0.18 & 15.08 & 17.82 & 0.15 & 15.07 & 18.61 & 0.19 \\
    NGC5897 & 2.62 & 6.17 & 0.57 & 2.55 & 6.22 & 0.59 & 2.62 & 6.21 & 0.58 & 2.57 & 6.30 & 0.59 \\
    NGC5904 & 2.57 & 18.96 & 0.86 & 2.95 & 19.69 & 0.85 & 2.99 & 19.28 & 0.84 & 3.56 & 19.15 & 0.81 \\
    NGC5927 & 4.12 & 4.75 & 0.13 & 4.16 & 4.78 & 0.13 & 4.07 & 4.74 & 0.14 & 4.11 & 4.76 & 0.14 \\
    NGC5946 & 0.54 & 3.31 & 0.84 & 0.49 & 3.31 & 0.85 & 0.51 & 3.29 & 0.84 & 0.52 & 3.30 & 0.84 \\
    BH176 & 12.50 & 18.89 & 0.34 & 11.99 & 19.00 & 0.37 & 12.35 & 18.40 & 0.33 & 12.73 & 18.76 & 0.32 \\
    NGC5986 & 0.50 & 2.76 & 0.82 & 0.43 & 2.74 & 0.84 & 0.51 & 2.77 & 0.82 & 0.46 & 2.74 & 0.83 \\
    FSR1716 & 4.14 & 6.31 & 0.34 & 4.20 & 6.37 & 0.34 & 4.17 & 6.22 & 0.33 & 4.20 & 6.28 & 0.33 \\
    Pal14 & 4.24 & 68.34 & 0.94 & 3.36 & 69.22 & 0.95 & 3.52 & 71.64 & 0.95 & 2.34 & 73.20 & 0.97 \\
    BH184 & 1.85 & 3.19 & 0.42 & 1.83 & 3.20 & 0.43 & 1.85 & 3.18 & 0.42 & 1.82 & 3.19 & 0.43 \\
    NGC6093 & 0.64 & 2.26 & 0.72 & 0.63 & 2.26 & 0.72 & 0.67 & 2.29 & 0.71 & 0.66 & 2.30 & 0.71 \\
    NGC6121 & 0.04 & 3.13 & 0.99 & 0.06 & 3.16 & 0.98 & 0.05 & 3.14 & 0.98 & 0.05 & 3.16 & 0.99 \\
    NGC6101 & 10.82 & 42.25 & 0.74 & 10.16 & 41.28 & 0.75 & 10.83 & 39.06 & 0.72 & 10.78 & 38.68 & 0.72 \\
    NGC6144 & 1.98 & 2.98 & 0.34 & 1.98 & 2.98 & 0.34 & 1.96 & 2.97 & 0.34 & 1.98 & 2.97 & 0.33 \\
    NGC6139 & 1.36 & 2.63 & 0.48 & 1.36 & 2.63 & 0.48 & 1.36 & 2.62 & 0.48 & 1.36 & 2.63 & 0.48 \\
    Terzan3 & 2.08 & 2.88 & 0.28 & 2.12 & 2.89 & 0.27 & 2.08 & 2.87 & 0.27 & 2.12 & 2.91 & 0.27 \\
    NGC6171 & 0.86 & 2.14 & 0.60 & 0.84 & 2.14 & 0.61 & 0.86 & 2.15 & 0.60 & 0.84 & 2.15 & 0.61 \\
    ESO45211 & 0.18 & 1.26 & 0.86 & 0.18 & 1.26 & 0.86 & 0.23 & 1.29 & 0.82 & 0.17 & 1.26 & 0.87 \\
    NGC6205 & 1.27 & 4.93 & 0.74 & 1.26 & 5.12 & 0.75 & 1.23 & 5.06 & 0.76 & 1.18 & 5.18 & 0.77 \\
    NGC6229 & 0.92 & 15.88 & 0.94 & 0.55 & 15.82 & 0.97 & 0.53 & 16.37 & 0.97 & 0.57 & 16.44 & 0.97 \\
    NGC6218 & 2.28 & 3.52 & 0.35 & 2.27 & 3.55 & 0.36 & 2.25 & 3.53 & 0.36 & 2.24 & 3.57 & 0.37 \\
    FSR1735 & 0.62 & 3.19 & 0.81 & 0.61 & 3.20 & 0.81 & 0.67 & 3.20 & 0.79 & 0.65 & 3.19 & 0.79 \\
    NGC6235 & 3.15 & 5.74 & 0.45 & 3.11 & 5.73 & 0.46 & 3.40 & 5.84 & 0.42 & 3.31 & 5.79 & 0.43 \\
    NGC6254 & 2.12 & 3.58 & 0.41 & 2.16 & 3.61 & 0.40 & 2.18 & 3.61 & 0.40 & 2.17 & 3.62 & 0.40 \\
    NGC6256 & 0.39 & 2.41 & 0.84 & 0.39 & 2.40 & 0.84 & 0.37 & 2.40 & 0.85 & 0.38 & 2.39 & 0.84 \\
    Pal15 & 1.45 & 29.16 & 0.95 & 1.39 & 31.10 & 0.96 & 2.27 & 32.01 & 0.93 & 1.27 & 33.69 & 0.96 \\
    NGC6266 & 0.95 & 1.43 & 0.33 & 0.94 & 1.43 & 0.34 & 0.95 & 1.43 & 0.34 & 0.94 & 1.43 & 0.34 \\
    NGC6273 & 1.03 & 3.26 & 0.69 & 0.98 & 3.24 & 0.70 & 0.90 & 3.21 & 0.72 & 0.89 & 3.21 & 0.72 \\
    NGC6284 & 0.42 & 4.20 & 0.90 & 0.45 & 4.22 & 0.89 & 0.42 & 4.24 & 0.90 & 0.39 & 4.24 & 0.91 \\
    NGC6287 & 0.45 & 3.94 & 0.89 & 0.54 & 4.06 & 0.87 & 0.49 & 3.96 & 0.88 & 0.43 & 4.00 & 0.89 \\
    NGC6293 & 0.50 & 2.07 & 0.76 & 0.48 & 2.06 & 0.77 & 0.51 & 2.08 & 0.75 & 0.50 & 2.07 & 0.76 \\
    NGC6304 & 1.70 & 2.35 & 0.28 & 1.70 & 2.36 & 0.28 & 1.69 & 2.35 & 0.28 & 1.70 & 2.36 & 0.28 \\
    NGC6316 & 0.76 & 2.12 & 0.64 & 0.76 & 2.11 & 0.64 & 0.76 & 2.12 & 0.64 & 0.76 & 2.11 & 0.64 \\
    NGC6341 & 0.36 & 5.47 & 0.93 & 0.14 & 5.39 & 0.97 & 0.17 & 5.55 & 0.97 & 0.29 & 5.64 & 0.95 \\
    NGC6325 & 0.36 & 1.46 & 0.75 & 0.29 & 1.41 & 0.80 & 0.31 & 1.43 & 0.78 & 0.32 & 1.43 & 0.77 \\
    NGC6333 & 0.92 & 4.84 & 0.81 & 0.96 & 4.83 & 0.80 & 1.04 & 4.94 & 0.79 & 1.05 & 4.93 & 0.79 \\
    NGC6342 & 0.97 & 1.60 & 0.39 & 0.96 & 1.59 & 0.40 & 0.97 & 1.60 & 0.39 & 0.97 & 1.60 & 0.39 \\
    NGC6356 & 3.39 & 5.96 & 0.43 & 3.30 & 5.86 & 0.44 & 3.28 & 5.99 & 0.45 & 3.16 & 5.87 & 0.46 \\
    NGC6355 & 0.50 & 2.04 & 0.75 & 0.48 & 2.03 & 0.76 & 0.52 & 2.04 & 0.75 & 0.50 & 2.03 & 0.75 \\
    NGC6352 & 3.05 & 3.48 & 0.12 & 3.05 & 3.50 & 0.13 & 3.01 & 3.47 & 0.13 & 3.02 & 3.49 & 0.14 \\
    IC1257 & 1.61 & 10.10 & 0.84 & 1.53 & 10.47 & 0.85 & 1.77 & 10.25 & 0.83 & 1.66 & 10.58 & 0.84 \\
    Terzan2 & 0.32 & 0.77 & 0.59 & 0.32 & 0.77 & 0.59 & 0.31 & 0.76 & 0.59 & 0.31 & 0.77 & 0.59 \\
    NGC6366 & 2.06 & 3.88 & 0.47 & 2.03 & 3.90 & 0.48 & 2.00 & 3.88 & 0.49 & 1.98 & 3.93 & 0.50 \\
    Terzan4 & 0.40 & 0.78 & 0.49 & 0.39 & 0.78 & 0.50 & 0.39 & 0.78 & 0.50 & 0.39 & 0.78 & 0.50 \\
    BH229 & 0.43 & 1.17 & 0.63 & 0.41 & 1.16 & 0.65 & 0.37 & 1.14 & 0.68 & 0.41 & 1.16 & 0.65 \\
    NGC6362 & 2.74 & 4.00 & 0.31 & 2.82 & 4.02 & 0.30 & 2.80 & 3.98 & 0.30 & 2.82 & 4.01 & 0.30 \\
    NGC6380 & 0.29 & 1.89 & 0.84 & 0.29 & 1.88 & 0.84 & 0.31 & 1.88 & 0.84 & 0.30 & 1.88 & 0.84 \\
    Terzan1 & 0.27 & 0.89 & 0.69 & 0.27 & 0.89 & 0.69 & 0.28 & 0.89 & 0.69 & 0.28 & 0.89 & 0.69 \\
    Ton2 & 0.90 & 2.38 & 0.62 & 0.91 & 2.39 & 0.62 & 0.90 & 2.37 & 0.62 & 0.91 & 2.39 & 0.62 \\
    NGC6388 & 1.18 & 2.26 & 0.48 & 1.18 & 2.26 & 0.48 & 1.19 & 2.26 & 0.47 & 1.19 & 2.26 & 0.47 \\
    NGC6402 & 0.70 & 2.39 & 0.71 & 0.69 & 2.39 & 0.71 & 0.69 & 2.42 & 0.72 & 0.70 & 2.43 & 0.71 \\
    NGC6401 & 2.66 & 4.02 & 0.34 & 2.62 & 4.00 & 0.35 & 2.72 & 4.06 & 0.33 & 2.65 & 4.03 & 0.34 \\
    NGC6397 & 2.68 & 4.50 & 0.40 & 2.57 & 4.55 & 0.43 & 2.67 & 4.49 & 0.41 & 2.62 & 4.56 & 0.43 \\
    Pal6 & 0.65 & 2.31 & 0.72 & 0.64 & 2.32 & 0.72 & 0.65 & 2.31 & 0.72 & 0.64 & 2.32 & 0.73 \\
    NGC6426 & 4.27 & 11.33 & 0.62 & 4.11 & 11.82 & 0.65 & 4.23 & 11.38 & 0.63 & 4.04 & 11.26 & 0.64 \\
    Djorg1 & 0.98 & 5.55 & 0.82 & 0.96 & 5.59 & 0.83 & 0.91 & 5.51 & 0.83 & 0.92 & 5.55 & 0.84 \\
    Terzan5 & 0.28 & 0.88 & 0.68 & 0.28 & 0.88 & 0.68 & 0.28 & 0.88 & 0.68 & 0.28 & 0.88 & 0.68 \\
    NGC6440 & 0.37 & 0.89 & 0.58 & 0.37 & 0.89 & 0.58 & 0.37 & 0.89 & 0.59 & 0.37 & 0.90 & 0.59 \\
    NGC6441 & 1.05 & 2.46 & 0.57 & 1.03 & 2.46 & 0.58 & 1.04 & 2.46 & 0.58 & 1.02 & 2.45 & 0.58 \\
    Terzan6 & 0.33 & 1.10 & 0.70 & 0.33 & 1.10 & 0.70 & 0.34 & 1.11 & 0.69 & 0.33 & 1.10 & 0.70 \\
    NGC6453 & 0.96 & 2.73 & 0.65 & 0.96 & 2.72 & 0.65 & 0.95 & 2.71 & 0.65 & 0.93 & 2.73 & 0.66 \\
    NGC6496 & 4.04 & 9.07 & 0.55 & 3.64 & 8.89 & 0.59 & 3.84 & 8.93 & 0.57 & 3.71 & 8.85 & 0.58 \\
    Terzan9 & 0.08 & 0.66 & 0.89 & 0.07 & 0.67 & 0.90 & 0.07 & 0.67 & 0.89 & 0.08 & 0.57 & 0.86 \\
    Djorg2 & 0.86 & 1.94 & 0.55 & 0.87 & 1.94 & 0.55 & 0.86 & 1.93 & 0.56 & 0.88 & 1.95 & 0.55 \\
    NGC6517 & 0.64 & 2.52 & 0.74 & 0.63 & 2.51 & 0.75 & 0.64 & 2.53 & 0.75 & 0.63 & 2.52 & 0.75 \\
    Terzan10 & 0.87 & 4.35 & 0.80 & 0.81 & 4.28 & 0.81 & 1.05 & 4.42 & 0.76 & 1.01 & 4.38 & 0.77 \\
    NGC6522 & 0.24 & 0.77 & 0.69 & 0.24 & 0.77 & 0.69 & 0.24 & 0.77 & 0.69 & 0.23 & 0.77 & 0.70 \\
    NGC6535 & 1.44 & 2.88 & 0.50 & 1.41 & 2.88 & 0.51 & 1.44 & 2.89 & 0.50 & 1.42 & 2.89 & 0.51 \\
    NGC6528 & 0.51 & 1.27 & 0.60 & 0.52 & 1.25 & 0.59 & 0.52 & 1.27 & 0.59 & 0.52 & 1.27 & 0.59 \\
    NGC6539 & 2.00 & 2.70 & 0.26 & 2.00 & 2.72 & 0.27 & 1.98 & 2.71 & 0.27 & 1.96 & 2.72 & 0.28 \\
    NGC6540 & 1.56 & 2.17 & 0.28 & 1.55 & 2.17 & 0.29 & 1.56 & 2.17 & 0.28 & 1.55 & 2.17 & 0.29 \\
    NGC6544 & 0.11 & 2.67 & 0.96 & 0.09 & 2.68 & 0.97 & 0.10 & 2.66 & 0.96 & 0.09 & 2.69 & 0.97 \\
    NGC6541 & 1.39 & 2.88 & 0.52 & 1.35 & 2.89 & 0.53 & 1.36 & 2.88 & 0.53 & 1.33 & 2.89 & 0.54 \\
    ESO28006 & 1.59 & 8.22 & 0.81 & 1.67 & 8.35 & 0.80 & 1.71 & 8.26 & 0.79 & 1.58 & 8.32 & 0.81 \\
    NGC6553 & 2.00 & 2.58 & 0.23 & 2.01 & 2.59 & 0.22 & 1.99 & 2.59 & 0.23 & 2.00 & 2.59 & 0.23 \\
    NGC6558 & 0.29 & 1.39 & 0.79 & 0.29 & 1.39 & 0.79 & 0.29 & 1.39 & 0.79 & 0.29 & 1.39 & 0.79 \\
    Pal7 & 3.37 & 4.71 & 0.28 & 3.37 & 4.75 & 0.29 & 3.44 & 4.76 & 0.28 & 3.42 & 4.77 & 0.28 \\
    Terzan12 & 2.01 & 3.07 & 0.35 & 2.01 & 3.09 & 0.35 & 2.02 & 3.07 & 0.34 & 2.04 & 3.09 & 0.34 \\
    NGC6569 & 2.53 & 2.98 & 0.15 & 2.46 & 2.95 & 0.17 & 2.44 & 3.06 & 0.20 & 2.44 & 2.95 & 0.18 \\
    BH261 & 1.33 & 2.10 & 0.37 & 1.34 & 2.10 & 0.36 & 1.33 & 2.09 & 0.37 & 1.35 & 2.11 & 0.36 \\
    NGC6584 & 2.19 & 15.15 & 0.86 & 2.46 & 15.79 & 0.84 & 2.61 & 14.88 & 0.82 & 2.61 & 14.43 & 0.82 \\
    NGC6624 & 0.42 & 0.93 & 0.55 & 0.42 & 0.93 & 0.54 & 0.42 & 0.93 & 0.54 & 0.43 & 0.93 & 0.54 \\
    NGC6626 & 0.49 & 1.74 & 0.72 & 0.50 & 1.74 & 0.72 & 0.50 & 1.75 & 0.71 & 0.51 & 1.75 & 0.71 \\
    NGC6638 & 0.13 & 1.21 & 0.89 & 0.13 & 1.20 & 0.89 & 0.12 & 1.20 & 0.90 & 0.11 & 1.19 & 0.91 \\
    NGC6637 & 1.08 & 1.46 & 0.26 & 1.07 & 1.46 & 0.27 & 1.07 & 1.46 & 0.27 & 1.07 & 1.46 & 0.27 \\
    NGC6642 & 0.16 & 1.29 & 0.87 & 0.15 & 1.28 & 0.88 & 0.16 & 1.29 & 0.87 & 0.16 & 1.28 & 0.88 \\
    NGC6652 & 0.50 & 2.21 & 0.78 & 0.51 & 2.21 & 0.77 & 0.47 & 2.19 & 0.79 & 0.45 & 2.19 & 0.79 \\
    NGC6656 & 2.90 & 6.79 & 0.57 & 2.93 & 6.89 & 0.58 & 2.86 & 6.82 & 0.58 & 2.88 & 6.87 & 0.58 \\
    Pal8 & 2.46 & 4.02 & 0.39 & 2.39 & 4.01 & 0.41 & 2.38 & 4.03 & 0.41 & 2.33 & 4.02 & 0.42 \\
    NGC6681 & 0.17 & 3.07 & 0.94 & 0.12 & 3.07 & 0.96 & 0.09 & 3.04 & 0.97 & 0.14 & 3.06 & 0.96 \\
    NGC6712 & 0.44 & 2.77 & 0.84 & 0.42 & 2.78 & 0.85 & 0.41 & 2.77 & 0.85 & 0.39 & 2.77 & 0.86 \\
    NGC6715 & 15.68 & 49.77 & 0.69 & 16.04 & 89.12 & 0.82 & 12.80 & 46.77 & 0.73 & 16.01 & 66.21 & 0.76 \\
    NGC6717 & 1.17 & 1.78 & 0.34 & 1.17 & 1.78 & 0.35 & 1.17 & 1.78 & 0.34 & 1.17 & 1.79 & 0.34 \\
    NGC6723 & 1.40 & 2.64 & 0.47 & 1.31 & 2.62 & 0.50 & 1.34 & 2.62 & 0.49 & 1.29 & 2.62 & 0.51 \\
    NGC6749 & 1.63 & 3.32 & 0.51 & 1.60 & 3.31 & 0.52 & 1.62 & 3.33 & 0.51 & 1.60 & 3.33 & 0.52 \\
    NGC6752 & 3.50 & 4.48 & 0.22 & 3.49 & 4.49 & 0.22 & 3.58 & 4.52 & 0.21 & 3.54 & 4.50 & 0.21 \\
    NGC6760 & 2.09 & 3.77 & 0.45 & 2.08 & 3.78 & 0.45 & 2.07 & 3.79 & 0.46 & 2.06 & 3.80 & 0.46 \\
    NGC6779 & 0.61 & 6.57 & 0.91 & 0.90 & 6.81 & 0.87 & 0.55 & 6.71 & 0.92 & 0.59 & 6.77 & 0.91 \\
    Terzan7 & 13.39 & 39.41 & 0.66 & 8.12 & 30.51 & 0.73 & 11.23 & 37.12 & 0.70 & 6.97 & 29.56 & 0.76 \\
    Pal10 & 4.02 & 5.52 & 0.27 & 3.96 & 5.55 & 0.29 & 4.08 & 5.55 & 0.26 & 4.02 & 5.56 & 0.28 \\
    Arp2 & 18.58 & 56.30 & 0.67 & 18.38 & 98.39 & 0.81 & 15.71 & 51.68 & 0.70 & 9.48 & 49.00 & 0.81 \\
    NGC6809 & 1.10 & 3.60 & 0.70 & 1.09 & 3.62 & 0.70 & 1.09 & 3.63 & 0.70 & 1.08 & 3.64 & 0.70 \\
    Terzan8 & 16.71 & 51.19 & 0.67 & 9.10 & 38.29 & 0.76 & 14.29 & 47.28 & 0.70 & 7.70 & 37.21 & 0.79 \\
    Pal11 & 4.35 & 6.33 & 0.31 & 4.20 & 6.25 & 0.33 & 4.37 & 6.42 & 0.32 & 4.19 & 6.30 & 0.33 \\
    NGC6838 & 4.98 & 6.02 & 0.17 & 4.90 & 6.09 & 0.20 & 4.92 & 6.05 & 0.19 & 4.87 & 6.13 & 0.20 \\
    NGC6864 & 1.15 & 9.36 & 0.88 & 0.99 & 9.37 & 0.89 & 0.94 & 9.23 & 0.90 & 0.81 & 9.16 & 0.91 \\
    NGC6934 & 2.98 & 35.10 & 0.92 & 3.44 & 34.49 & 0.90 & 1.86 & 32.90 & 0.94 & 1.96 & 32.04 & 0.94 \\
    NGC6981 & 0.71 & 14.19 & 0.95 & 0.43 & 15.03 & 0.97 & 0.22 & 13.84 & 0.98 & 0.66 & 13.84 & 0.95 \\
    NGC7006 & 2.82 & 33.25 & 0.92 & 2.73 & 36.64 & 0.93 & 6.13 & 34.22 & 0.82 & 4.49 & 33.79 & 0.87 \\
    NGC7078 & 3.89 & 7.20 & 0.46 & 3.45 & 6.99 & 0.51 & 4.04 & 7.34 & 0.45 & 3.57 & 7.25 & 0.51 \\
    NGC7089 & 0.78 & 10.98 & 0.93 & 0.56 & 10.79 & 0.95 & 0.87 & 11.24 & 0.92 & 0.83 & 11.25 & 0.93 \\
    NGC7099 & 1.40 & 4.96 & 0.72 & 1.51 & 5.08 & 0.70 & 1.39 & 4.94 & 0.72 & 1.55 & 5.03 & 0.69 \\
    Pal12 & 15.63 & 66.42 & 0.76 & 8.76 & 52.09 & 0.83 & 13.10 & 59.79 & 0.78 & 13.39 & 49.53 & 0.73 \\
    Pal13 & 9.23 & 52.76 & 0.83 & 8.79 & 60.90 & 0.86 & 8.95 & 50.28 & 0.82 & 9.40 & 49.34 & 0.81 \\
    NGC7492 & 4.72 & 17.13 & 0.72 & 3.63 & 16.75 & 0.78 & 3.03 & 16.22 & 0.81 & 2.81 & 16.04 & 0.82 \\
\bottomrule
\caption{\emph{The pericentre, semi-major axis, and orbital eccentricity of each Galactic globular cluster when its orbit is integrated in the potential of the Milky Way (MW), the combined potential of the Milky Way and the Sagittarius Dwarf Galaxy (SGR), the combined potential of the Milky Way and the Magellanic Clouds (SMC/LMC), and the combined potential of the Milky Way and the all six dwarf galaxies considered. While not listed, clusters have been sorted by right ascension to be consistent with \citet{Harris96}.}}
\label{orbits}
\end{longtable}
\end{center}


\label{lastpage}
\end{document}